\documentclass[preprint]{JHEP3} 

\newcommand\fverb{\setbox\pippobox=\hbox\bgroup\verb}
\newcommand\fverbdo{\egroup\medskip\noindent%
            \fbox{\unhbox\pippobox}\ }
\newcommand\fverbit{\egroup\item[\fbox{\unhbox\pippobox}]}
\newbox\pippobox

\newcommand{\be}{\begin{equation}}
\newcommand{\ee}{\end{equation}}

\newcommand{\bea}{\begin{eqnarray}}
\newcommand{\eea}{\end{eqnarray}}

\newcommand{\eda}{\eta^\dagger}

\title{Comments on $AdS_{2}$ solutions of $D=11$ Supergravity}

\author{Nakwoo Kim and Jong-Dae Park\\
    Department of Physics and Research Institute of Basic Science,\\
    Kyung Hee University, Seoul 130-701, Korea \\
    E-mail: \email{nkim@khu.ac.kr}, \email{jdpark@khu.ac.kr}}

\abstract{
We study the supersymmetric solutions of 11-dimensional supergravity 
with a factor of $AdS_2$ made of M2-branes.
Such solutions can provide gravity duals of superconformal
quantum mechanics, or through double Wick rotation, the generic bubbling geometry of M-theory which are 1/16-BPS. We show that, when the internal
manifold is compact, it should take the form of a warped U(1)-fibration over an 
8-dimensional K\"ahler space. 
}


\begin{document}


\section{Introduction}
It is one of the most intriguing issues in string theory to prove/disprove 
the so-called Maldacena conjecture between anti-de-Sitter gravity and 
conformal field theories \cite{AdSCFT}. A lot of remarkable agreements have been 
encountered so far, so it would be more appropriate to say one would like
to determine the region of validity as precise as possible. 

This inevitably
leads us to the study of {\it less} supersymmetric solutions of String/M-theory. 
The spectrum of supersymmetric solutions is rich enough to provide 
examples with realistic features such as confinement and asymptotic 
freedom (see \cite{confining} for example), 
and yet thanks to the power of supersymmetry we can perform
explicit checks using protected quantities. 

In this paper we will employ a technique which is found to be 
very powerful and at the same time illuminating in the search for 
new supersymmetric backgrounds of String/M-theory. Based on 
the existence of Killing spinors, one can construct various differential
forms as spinor bilinears and determine the local form of the metric
and the gauge fields exploiting the differential and algebraic relations
between the differential forms which can be derived using the
Killing spinor equations. A sample of works which make use of this
technique can be found 
in Ref.\cite{KStech}. 

In this paper, as a sequel to the previous one \cite{Nkim05}, 
we analyze the consequences of supersymmetry in 11-dimensional 
supergravity, combined with the ansatz of $AdS_2$ factor, i.e. 
$SO(2,1)$ isometry. In particular, for simplicity, we consider 
pure M2-brane configurations. A convenient way of seeing them 
is as M2-branes wrapped on two-cycles in a Calabi-Yau 
5-fold, and the readers are referred to \cite{m2} for studies along 
alternative avenues on similar configurations. In the present work, 
it will be shown that, the 9-dimensional internal manifold should
take the form of a warped U(1)-fibration on a K\"ahler base. We also 
write down the 8-dimensional nonlinear partial differential equation 
for the curvature tensor, which is required if the supersymmetric
configuration is to solve the equations of motion. 

This paper is organised as follows. In Section 2, we set up 
the problem and derive the 9-dimensional Killing equations
from the 11-dimensional one. In Section 3, we consider the 
spinor bilinears and their derivatives to fix the local form of
the metric. In Section 4 we illustrate that the well-known solutions
which have $AdS_2$ factors can be indeed cast in the form 
as we have presented in this paper. In Section 5 we conclude with 
comments and discussions on further works.

\section{Ansatz}
In this article we consider supersymmetric solutions of D=11 supergravity, whose
Lagrangian density in the bosonic sector is given as follows,
\begin{eqnarray}
{\cal L} =    R \, *1 - \frac{1}{2}
G \wedge * G  - \frac{1}{6} C \wedge
G \wedge G , 
\end{eqnarray}
where $C$ is the 3-form potential and $G=dC$. 

By definition, supersymmetric backgrounds allow nontrivial solutions
to the Killing spinor equation which is obtained by setting the supersymmetry
transformation to zero. For D=11 supergravity, we thus have, for purely bosonic
backgrounds,
\begin{eqnarray}
\delta\psi_{M} = \nabla_{M} \epsilon +
\frac{1}{288}\left(\Gamma_{M}^{~M_1 \cdots
M_4}-8\delta_{M}^{~M_1}\Gamma^{M_2 M_3 M_4}\right)G_{M_1\cdots
M_4}\epsilon = 0 , 
\end{eqnarray}
where $\epsilon$ is the supersymmetry parameter which is a Majorana 
spinor in $D=11$ and  $\Gamma_M , \,\, M=0,1,\cdots , 10$ are the
gamma matrices satisfying $\{\Gamma_M, \Gamma_N\} = 2 g_{MN}$. 

Now we restrict ourselves to a specific class of solutions which have 
an $AdS_2$ factor. We can write the metric as follows,
\be
ds^2 = e^{2A} ds^2 (AdS_2)+g_{ab}dx^a dx^b, ~~~a,b = 1,2,\cdots,9 \, . 
\ee
$A$ is the warp factor and a function of $x^a$ only, and it does not depend 
on the coordinates of $AdS_2$. Technically we are performing a dimensional
reduction on $AdS_2$ to get a 9 dimensional system from the 11 dimensional
supergravity, but the Euclidean space defined by $x^a$ will be referred to as  
{\it internal}. Without loss generality we set the radius
of $AdS_2$ to 1, i.e. the scalar curvature is 2. 

Furthermore, we consider {\it electric} ansatz
for the field strength $G$, and take
\be
G = {\rm Vol}_{AdS_2} \wedge F . 
\ee
Note that most generally one can also consider turning on the 4-form fields in 
the internal 9 dimensional space. Our setup amounts to the 
consideration of pure M2-brane configurations, and among other things trivializes the
effect of  
the cubic interaction of the gauge field $C$. The equation of motion and the 
Bianchi identity for $C$ are reduced to the following expressions for the 
two-form field $F$ in 
the 9 dimensional internal space, 
\begin{eqnarray}
&& d(e^{-2A}*F)=0,~~~ dF=0.
\end{eqnarray}

We now need to introduce the basis for the gamma matrices which is 
convenient for 
the dimensional decomposition we have chosen to consider. With
tangent space indices, they are 
\begin{eqnarray}
\Gamma_{\mu} &=& \gamma_{\mu} \otimes 1 , ~~~ \mu = 0, 1,  \\
\Gamma_a &=& \gamma_3 \otimes \gamma_a, ~~~ a = 2, \cdots ,10 , 
\end{eqnarray}
where $\gamma_\mu$'s are the 2 dimensional gamma matrices, while 
$\gamma_a$'s denote 9 dimensional ones. For simplicity we take the basis
where all $\gamma_\mu$ and $\gamma_a$ are real, and of course $\Gamma_M$'s
are also real matrices. 
 
 As the ansatz for the Killing spinor $\epsilon$, we assume its 2 
 dimensional part satisfies
\begin{eqnarray}
\label{onads2}
\bar{\nabla}_{\mu}\varepsilon =
\frac{a}{2}\gamma_{\mu}\varepsilon,~~~a=\pm 1 . 
\end{eqnarray}
Here $\bar{\nabla}$ denotes the covariant derivative on a unit-radius
$AdS_2$. There is an alternative to this equation, i.e. $\bar{\nabla}_\mu 
\varepsilon = \frac{ai}{2}\gamma_\mu \gamma_3 \varepsilon$, but in 2 
dimensions this just corresponds to the parity inversion so for definiteness
we choose Eq.(\ref{onads2}). We can then take the 11 dimensional 
spinor $\epsilon = \varepsilon \otimes \eta + c.c. $, where $\eta$ is 
a 9 dimensional spinor.  

Combining the ingredients given above, it is straightforward to derive
the 9 dimensional Killing spinor equations, which can be presented as 
follows. 
\begin{eqnarray}
&& a i e^{-A}\eta + /\!\!\!\partial
A\eta-\frac{1}{6}
e^{-2A}/\!\!\!\! F\eta=0\label{sb1} , \\
&&\nabla_{a}\eta+\frac{1}{24}e^{-2A}\left(\gamma_{a}^{~bc}F_{bc}
-4F_{ab}\gamma^{b}\right) \eta=0 . \label{sb2}
\end{eqnarray}
In the following we will analyze how the above equations restrict the
local form of the 9-dimensional internal metric. 
%
%
\section{Spinor bilinears and the consequences of supersymmetry}
We now consider the various spinor bilinears made out of $\eta$, and exploit the
Killing spinor equations to find the local form of the metric. We can consider 
the real-valued differential forms such as 
\bea
f &=& \eda \eta ,\\
K &=& \eda \gamma_a \eta \, dx^a ,\\
Y &=& \frac{ai}{2} \eda \gamma_{ab} \eta \,\,  dx^a \wedge dx^b  . 
\eea
For instance, with the zero-form $f$ one can easily verify that, using
Eq.(\ref{sb1}) and Eq.(\ref{sb2}), 
\begin{eqnarray}
\nabla_{a}(\eta^{\dagger}\eta) 
&=& \frac{1}{3}e^{-2A}F_{ac} 
\eta^{\dagger}\gamma^{c}\eta \label{1f} \\
&=& \partial_{a}A \eta^{\dagger}\eta , 
\end{eqnarray}
implying that one can set $\eta^{\dag}\eta=e^{A}$. We proceed in the
same way and find that $\nabla_a K_b + \nabla_b K_a = 0 $, i.e. $K$ 
defines a Killing vector in the internal space, and 
\be
d ( e^A K ) = F + Y . 
\label{2f}
\ee
For the two-form $Y$, we get simply
\be 
dY = 0 . 
\label{3f}
\ee 

Now we can choose the coordinate system where
$K = \partial_\psi$ and the metric of the internal space is 
\be
ds^2 = e^{2\phi} (d\psi + B)^2 + g_{ij} dy^i dy^j \quad i,j = 1,2, \cdots ,8. 
\ee
$\phi, B$ are respectively a scalar and a vector field defined on the 8 dimensional
space ${\cal M}_8$ with coordinates $y^i$. 

Now we claim that, when the 9 dimensional internal space is compact, $\eta$
is a chiral spinor on ${\cal M}$, and as a result $\phi=A$. We will need to 
make use of the following identities which hold for an arbitrary Dirac spinor 
$\eta$ in 
9-dimensions.
\bea
(\eta^T \eta )^2 &=& ( \eta^T \gamma_a \eta )^2 , \label{fi1} \\
|\eta^T\eta| - | \eta^T\gamma_a \eta|^2  &=& 2 (\eda \gamma_a \eta )^2
- 2 (\eda \eta)^2 \label{fi2},
\eea
whose derivation can be found in Ref. \cite{Kim:2006ag}. 

We consider the spinor bilinear $\eta^T\eta$, which is complex-valued. 
From the Killing equations we get
\be
\partial_a ( e^{-A} \eta^T \eta  ) = a i e^{-2A} \, \eta^T \gamma_a \eta . 
\ee
For $D=e^{-A}\eta^T\eta$, we thus have $(\nabla_a D)^2  = - e^{-4A} 
(\eta^T\gamma_a \eta)^2 = - e^{-2A} D^2 $.  One can also show that, from
the Killing equations Eq.(\ref{sb1}) and Eq.(\ref{sb2}), 
\bea
\nabla^2 ( e^{-A} \eta^T\eta ) 
&=& ai \nabla^a ( e^{-2A} \eta^T \gamma_a \eta )
\nonumber \\
&=&
-2 e^{-3A} \eta^T \eta , 
\eea
i.e. $\nabla^2 D + 2 e^{-2A} D = 0 $. From these relations, it is easy to
see that $\nabla^2 (D^{-1}) = 0$ provided $D$ is not zero. 
If the internal space is compact, this is
possible only if $D$ is constant, but it means $D=0$, so we have 
a contradiction. We thus conclude $\eta^T\eta=\eta^T\gamma_a\eta=0$,
and from Eq.(\ref{fi2}) $\eta$ is chiral on ${\cal M}_8$ 
and we have $K^2 = e^{2\phi} = e^{2A}$. 

Obviously the above argument requires the internal space should be 
compact, and 
$A$ should not show a singular behavior. But in the next section we will 
show that, for the important class of solutions such as $AdS_4\times SE_7$
and the bubbling geometry of Ref. \cite{Lin:2004nb}, even though the 
internal manifolds are not compact, the solutions can be rewritten in the 
manner we will conclude in this section. We guess that it might be
possible to improve our proof for the chirality of $\eta$, without
assuming compact internal space. 

Now that $\eta$ is chiral, $Y$ is a closed two-form in ${\cal M}_8$, which can
be used to define an almost complex structure. In order to see whether
this complex structure is integrable or not, one needs to check the 
exterior derivative of the 
complex-valued $(4,0)$-form $\Omega$, defined as 
\be
\Omega_{abcd} = \eta^T \gamma_{abcd} \eta . 
\ee
The chirality of $\eta$ again restricts $\Omega$ to be a four-form on ${\cal M}_8$,
and as usual with $SU(n)$-structures, $J,\Omega$ satisfy
\be
{\rm Vol}_{{\cal M}_8 }= \frac{e^{-4A}}{24} J\wedge J\wedge J\wedge J 
= \frac{e^{-2A}}{16} \Omega \wedge \bar{\Omega} , 
\quad J \wedge \Omega = 0 . 
\ee
Using the Killing equations one obtains
\be
d ( e^A \Omega ) = -ai e^{-A} K \wedge \Omega . 
\ee
For $\omega \equiv e^{A}e^{-ia\psi}\Omega$ and using $K = e^{2A} (d\psi + B)$, 
we have 
\begin{eqnarray}
d\omega = -ai B \wedge \omega , 
\end{eqnarray}
which is an 8 dimensional equation on $ {\cal M}_8 $. From the general
result of $SU(n)$-structures and the  classification of torsion classes, we arrive
at the conclusion that the complex structure given by $Y$ is integrable and
$dB={\cal R}$ is the Ricci form of the K\"{a}hler
manifold ${\cal M}_8$. Considering $Y^2 \sim (\eda\eta)^2 \sim e^{2A}$ and
$\omega^2 \sim e^{2A} |\eta^T\eta|^2 \sim e^{4A}$ when evaluated using the 
metric $g$, 
it is the rescaled metric $\bar{g}_{ij} = e^{A} g_{ij}$ which is K\"ahler. 

One can rephrase Eq.(\ref{2f}) to get  
\bea
F &=& \bar{F} + 3e^{A}dA \wedge K , \\
e^{3A}{\cal R} &=& \bar{F} + Y  , 
\eea
%
%
where $\bar{F}$ is the two-form field $F$ restricted to ${\cal M}_8$. 
Now if we contract Eq.(\ref{sb1}) with $\eda$, we have 
$\bar{F}_{ij} Y^{ij} = - 6 e^{-2A}$. We thus have the expression for the
scalar curvature $R$ of ${\cal M}_8$ whose K\"ahler metric is given 
as $\bar{g}$, 
\be
R = 2 e^{-3A} . 
\ee

Consideration of other spinor bilinears does not generate independent 
equations, but we need to impose the Bianchi identity and the equation of 
motion for $F$, to guarantee that the supersymmetric configuration really
satisfies all the equations of motion \cite{Gauntlett:2002fz}. 
It turns out that $dF=0$ is a 
consequence of supersymmetry, as can be easily seen from 
Eq.(\ref{2f}) and Eq.(\ref{3f}). Using the equation of motion for $F$,
$\nabla^a ( e^{-2A} F_{ab} ) = 0 $, and Eq.(\ref{1f}), one can derive 
the following equation for the scalar curvature of ${\cal M}_8$. 
\begin{eqnarray}
\Box R - \frac{1}{2}R^2 + R_{ij}R^{ij} = 0.
\label{master}
\end{eqnarray}
We conclude that, our ansatz of supersymmetric $AdS_2$ requires the
metric should locally be written as 
\be
ds^2 = e^{2A} ds^2_{AdS_2} + e^{2A} (d\psi +B)^2 + e^{-A} \bar{g}_{ij} dy^i dy^j
,
\label{11d}
\ee
where $\bar{g}$ defines a K\"ahler metric, and $dB$ is the Ricci form of
the 8 dimensional base space with coordinates $y^i$.


\section{Examples of $AdS_2$ solutions and their K\"ahler geometry}

In this section we will illustrate that several classes of well-known solutions which include an
$AdS_2$ factor can be indeed written in the form given as Eq.(\ref{11d}). 

The simplest case is obtained when we take ${\cal M}_8 = S^2\times T^6$. 
Being related to the curvature scalar of ${\cal M}_8$,
$A$ is a constant, and the eleven dimensional spacetime becomes
$AdS_2 \times S^3 \times T^6$, where the radius of $S^3$ is twice as large
as that of $AdS_2$. This  solution is a well-known example, which is given
as the near-horizon limit of three M2-branes intersecting over a point. As such,
the preserved supersymmetry is in fact 1/4 for this solution.

We next consider $AdS_{4} \times SE_7$ solutions where $SE_7$ is a
 7 dimensional
Sasaki-Einstein manifold. They can be considered as the near-horizon limit of
M2-brane solutions when put on a singularity of Calabi-Yau 4-folds. As such, 
in general these solutions are 1/8-BPS.  

It is well
known that in canonical form, any Sasaki-Einstein manifolds can be written as a
U(1)-fibration over a K\"ahler-Einstein manifold. For $SE_7$, using the standard
convention, 
\be
ds^2 = (d\alpha + \frac{\sigma}{4})^2 + ds^2_{KE_6}, 
\ee
where $d\sigma = {\cal R} = 8 J$. $R$ is the Ricci form and $J$ is the
K\"ahler form of the 6 dimensional K\"ahler-Einstein manifold. In the following 
we will rewrite the 11-dimensional metric in a form where the 8-dimensional
K\"ahler structure is manifest.
\begin{eqnarray}
ds^2 &=& \frac{1}{4}ds^2_{AdS_4} + ds^2_{SE_7}  \nonumber\\
&=& \frac{1}{4}\left[\cosh^{2}\rho \,ds^2_{AdS_2} + d\rho^2 + \sinh^{2}\rho
d\phi^2 \right]+ \left(d\alpha + \frac{\sigma}{4} \right)^2 + ds^2_{KE_6}
\nonumber\\
&=& \frac{1}{4} \cosh^{2}\rho  \,ds^2_{AdS_2} + \frac{1}{4} \cosh^{2}\rho
\left(d\phi + 2\frac{d\tilde{\alpha} + \sigma/4}{\cosh^{2}\rho}
\right)^2  \nonumber\\
&& + \frac{2}{\cosh\rho} \left( \frac{\cosh\rho}{2} (\frac{d\rho^2}{4} + ds^2_{KE_6}) +  \frac{\sinh^{2}\rho}{2\cosh\rho}
\left( d\tilde{\alpha} + \frac{\sigma}{4} \right)^2 \right), 
\end{eqnarray}
where we put $\alpha \rightarrow \tilde{\alpha} + \phi/2$.\\

Compared to the general form of the solution given in Eq.(\ref{11d}), 
evidently we expect that the 8-dimensional metric 
\be
ds^2 = \frac{\cosh\rho}{2} \left(
\frac{d\rho^2}{4} + ds^2_{KE_6}
\right) +  \frac{\sinh^{2}\rho}{2\cosh\rho}
\left( d\tilde{\alpha} + \frac{\sigma}{4} \right)^2 
\ee
should be K\"ahler, and the Ricci-form is given as
\be
{\cal R} =  d 
\left( 2 
\frac{d\tilde{\alpha} + \sigma/4}{\cosh^{2}\rho}
\right) , 
\label{ricci}
\ee
with the scalar curvature 
\be
R = \frac{16}{\cosh^3\rho} , 
\ee
and finally, Eq.(\ref{master}) should be satisfied. 

In order to check this, it is most efficient to construct the almost K\"ahler
form $J$ and the $(4,0)$-form $\Omega$, and compute their exterior derivatives.
A reasonable guess for $J$ is
\begin{eqnarray}
J = \frac{1}{4}\sinh\rho d\rho \wedge \left( d\tilde{\alpha} +
\frac{\sigma}{4} \right) + \frac{\cosh\rho}{2} J_{KE_6} . 
\end{eqnarray}
One can readily check that $dJ=0$. From the complex structure given by $J$, 
the (4,0)-form $\Omega$ is given as 
\begin{eqnarray}
\Omega &=& \frac{\cosh\rho}{8} \left( \cosh\rho d\rho + 2i\sinh\rho \left(
d\tilde{\alpha} + \frac{\sigma}{3} \right) \right) \wedge
\Omega_{KE_6} . 
\end{eqnarray}
One can also check that, as required by Eq.(\ref{ricci}), 
\be
d\Omega = \frac{2i}{\cosh^2\rho} (d\tilde{\alpha}+\frac{\sigma}{4})  \wedge \Omega . 
\ee 
And Eq.(\ref{master}) is indeed satisfied. 
%

The next example is the 1/2-BPS bubbling geometry of M-theory giant gravitons obtained in \cite{Lin:2004nb}. This class of solutions describe generic 1/2-BPS 
operators of M2-brane or M5-brane conformal field theories. From superalgebra
arguments, such configurations should possess $SO(3)\times SO(6)$ symmetry,
which are realized as a $S^2\times S^5$ factor in the metric. $S^2$ can be treated as a Wick-rotated $AdS_2$ to fit into our result.  Another comment in order is
that now the canonical Killing vector made out of the Killing spinor becomes 
time-like, so our solutions can describe magnetic, or M5-branes, as well as 
electric or M2-brane configurations. The solutions are summarized
as follows. 
\begin{eqnarray}
ds_{11}^{2}&=& -4e^{2\lambda}(1+y^{2}e^{-6\lambda})
(dt+V_{i}dx^{i})^2 +\frac{e^{-4\lambda}}
{1+y^{2}e^{-6\lambda}}[dy^{2} + e^{D}(dx_{1}^2+dx_{2}^{2})] \nonumber\\
&& + 4e^{2\lambda}d\Omega_{5}^{2} + y^{2}e^{-4\lambda}d{\tilde
\Omega}_{2}^{2}  \nonumber\\
G &=& F \wedge d\tilde{\Omega}^2_2 \nonumber \\
e^{-6\lambda} &=& \frac{\partial_y D}{y(1-y\partial_y D)} \nonumber\\
V_i &=& \frac{1}{2}\epsilon_{ij}\partial_{j}D ~~{\rm or}~~dV =
\frac{1}{2}*_3 [d(\partial_y D) + (\partial_y D)^2 dy] \nonumber\\
F &=& dB_t \wedge (dt+V) + B_t dV + d\hat{B} \nonumber\\
B_t &=& -4y^3 e^{-6\lambda} \nonumber\\
d\hat{B} &=& 2*_3 [(y\partial^2_y D + y(\partial_y D)^2 -
\partial_y D)dy + y\partial_i \partial_y D dx^i] \nonumber\\
&=& 2\tilde{*_3}[y^2(\partial_y \frac{1}{y}\partial_y e^D)dy + ydx^i
\partial_i \partial_y D]\label{LLM1}
\end{eqnarray}
where $i,j = 1,2$ and $*_3$ is the three dimensional $\epsilon$
symbol of the metric $dy^2 + e^D dx^2_i$ and $\tilde{*}_3$ is the
flat space $\epsilon$ symbol. The function $D$ satisfies the
equation
\begin{eqnarray}
(\partial^2_1 + \partial^2_2)D + \partial^2_y e^D = 0 . 
\end{eqnarray}

In order to see the K\"{a}hler structure the metric of the equation let us first 
rewrite (\ref{LLM1}) as follows,
\begin{eqnarray}
ds_{11}^{2} &=& y^{2}e^{-4\lambda}d{\tilde \Omega}_{2}^{2} +
4e^{2\lambda}
[(d\alpha + \frac{\sigma}{3})^2 + ds^2_{KE_4}]\nonumber\\
&&- 4e^{2\lambda}(1+y^2e^{-6\lambda})
(dt + V)^2\nonumber\\
&& + \frac{e^{-4\lambda}}{1 + y^2e^{-6\lambda}}[dy^{2} +
e^{D}(dx_{1}^2 +dx_{2}^{2})], 
\end{eqnarray}
where the 4-dimensional K\"ahler-Einstein manifold is $CP^2$, which 
provides the K\"ahler-Einstein base for the trivial example of 5-dimensional 
Sasaki-Einstein manifold $S^5$. In fact, as far as supersymmetry is 
concerned, any 4-dimensional K\"ahler-Einstein manifold would suffice
as long as it satisfies ${\cal R}=6J=d\sigma$. 

If we define $\alpha = {\tilde \alpha}+t$, then the metric
can be rewritten as
\begin{eqnarray}
ds_{11}^2 &=& y^2 e^{-4\lambda}d{\tilde \Omega}_2^2 -  4y^2
e^{-4\lambda}[dt-\frac{e^{6\lambda}}{y^2}
(d{\tilde \alpha}+\frac{\sigma}{3} - (1+y^2 e^{-6\lambda})V)]^2
\nonumber\\
&& +\frac{4}{ye^{-2\lambda}}\left\{ \frac{e^{6\lambda}}{y}
(1 + y^2e^{-6\lambda})(d{\tilde \alpha} + \frac{\sigma}{3} - V)^2\right.
\nonumber\\
&& \left. + \frac{1}{4}\frac{ye^{-6\lambda}}{1+y^2 e^{-6\lambda}}
[dy^2 + e^{D}(dx_1^2 + dx_2^2)] + yds_{KE_4}^2 \right\} . 
\end{eqnarray}
Now we are ready to identify the K\"ahler structure of the M-theory
bubbling solution. The K\"{a}hler form of the 8-dimensional base space is given by
\be
J = -\frac{1}{4} \partial_y (e^D) dx_1\wedge dx_2 + \frac{1}{2}dy\wedge
(d{\tilde \alpha} + \frac{\sigma}{3} - V) + yJ_{KE_4} , 
\ee
where $J_{KE_4}$ is the K\"{a}hler form of $KE_4$. 
One can easily show that $J$ is closed, and for the $(4,0)$-form 
given as 
\begin{eqnarray}
\Omega &=& 4 y e^{D/2}  \left( \partial_y D \, dy +
2i (d{\tilde \alpha} + \frac{\sigma}{3} - V)\right)
\wedge ( dx_2 + i dx_1 ) 
\wedge y\Omega_{KE_4} , 
\end{eqnarray}
and also we indeed have, as expected from the twisting of the U(1)-fibration 
in the metric, 
\be
d\Omega = 2i \left( \frac{e^{6\lambda}}{y^2} (d\tilde{\alpha} +\frac{\sigma}{3} )
- \frac{e^{6\lambda} + y^2}{y^2} V \right) \wedge \Omega . 
\ee
\section{Discussions}
In this work we have studied supersymmetric M2-brane configurations which have 
a factor of $AdS_2$. 
They can be interpreted as the near-horizon limit of M2-branes,
whose worldvolume is wrapped on a 2-cycle in a Calabi-Yau 5-fold. In general,
they are thus 1/16-BPS. It turns out that 
the internal 9-dimensional manifold
should take a form of U(1)-fibration whose base manifold is given by
a K\"ahler manifold. There is a restriction on the K\"ahler base imposed by supersymmetry: we have found a Laplace-like equation for the scalar 
curvature and Ricci tensor, as given in Eq.(\ref{master}).

The result
presented in this article is amusingly
very similar to the result of \cite{Nkim05}, where pure D3-brane configurations
with $AdS_3$ are studied. In that case, the internal manifold is 7-dimensional, which again
takes the form of warped U(1)-fibration on a 6-dimensional K\"ahler manifold.
The K\"ahler base cannot be arbitrary, it has to satisfy a nonlinear
partial differential equation for the curvature.

It is certainly of great interest to find new AdS solutions, by directly 
trying to solve Eq.(\ref{master}). In fact, recently a new class of $AdS_3$ 
solutions in IIB supergravity has been presented \cite{Gauntlett:2006af}. 
The authors first studied $AdS_3$ solutions of 11-dimensional supergravity, 
and then, through T-duality operations, obtained $AdS_3$ solutions of IIB supergravity, where only the
five-form fluxes are turned on. This is of course
very similar to the discovery of the celebrated Sasaki-Einstein solutions
$Y^{p,q}$ as part of $AdS_5$ solutions in IIB supergravity
\cite{Gauntlett:2004zh}. 
 In particular, it was argued that the new solutions can be 
indeed written in the form as presented in \cite{Nkim05}. 
The relevant 6-dimensional K\"ahler manifold takes a form of $S^2$-bundle over
a 4-dimensional K\"ahler-Einstein manifold. It will be very interesting to introduce
such a concrete ansatz, and solve Eq.(\ref{master}). We expect, as is the
case with $Y^{p,q}$, the metric of the K\"ahler base might be in general 
{\it not} complete, but the entire 9-dimensional internal manifold can be made 
complete. We plan to present the new solutions and the global analysis
in a future publication. 

Another interesting direction is to interpret our solutions as generalized 
bubbling geometry. For the case of $AdS_5\times S^5$ solutions in 
IIB supergravity, the 1/2-BPS bubbling solutions given in \cite{Lin:2004nb} 
can be described in terms of a distribution function in the phase space of
1-dimensional free fermions. For M-theory, determining the solutions is 
instead reduced to solving a Toda equation in 3-dimensions, and we believe
the dynamics of 1/2-BPS operators must be encoded therein. 
The supergravity solutions dual to less supersymmetric giant graviton 
operators are considered for 1/8-BPS in Ref.\cite{Nkim05} and for 1/4-BPS
in Ref.\cite{Donos:2006iy}. They both conclude that the 10 dimensional
solution is based on a K\"ahler space which is 6 and 4-dimensional, respectively. 
Naturally one expects they originate from the symplectic structure of the
eigenvalue dynamics. 
Likewise, 
Eq.(\ref{master}) can be interpreted as the equation governing the
dynamics of generic supersymmetric operators of M-branes conformal
field theory. One important feature we have been ignoring in this paper is
the global property and the boundary conditions of the solutions. 
Since our analysis is general enough to encompass the 1/2-BPS fluctuations 
of less supersymmetric conformal field theories on M-branes, as well as 
less supersymmetric fluctuations of maximal conformal field theories on 
M-branes, we will first need to fix the boundary condition 
according to the conformal field theory we are interested in, and then 
solve Eq.(\ref{master}) to find gravity duals to BPS operators. 

\acknowledgments
We are grateful to Ho-Ung Yee and Sang-Heon Yi for discussions. 
The research of Nakwoo Kim is supported by the Science Research Center
Program of the Korea Science and Engineering Foundation (KOSEF) through
the Center for Quantum Spacetime (CQUeST) of Sogang University with 
grant number R11-2005-021, and by the Basic Research Program of 
KOSEF with grant No. R01-2004-000-10651-0. Nakwoo Kim and Jong-Dae 
Park are both supported by the Korea Research Foundation Grant KRF-2003-070-C00011.





\begin{thebibliography}{999}

\bibitem{AdSCFT}
  J.~M.~Maldacena,
 {\em  The large N limit of superconformal field theories and supergravity,}
  Adv.\ Theor.\ Math.\ Phys.\  {\bf 2} (1998) 231
  [Int.\ J.\ Theor.\ Phys.\  {\bf 38} (1999) 1113]
  [arXiv:hep-th/9711200].
  
  S.~S.~Gubser, I.~R.~Klebanov and A.~M.~Polyakov,
{\em Gauge theory correlators from non-critical string theory},
  Phys.\ Lett.\ B {\bf 428} (1998) 105
  [arXiv:hep-th/9802109].
  
  E.~Witten,
{\em Anti-de Sitter space and holography},
  Adv.\ Theor.\ Math.\ Phys.\  {\bf 2} (1998) 253
  [arXiv:hep-th/9802150].
  
  O.~Aharony, S.~S.~Gubser, J.~M.~Maldacena, H.~Ooguri and Y.~Oz,
  {\em Large N field theories, string theory and gravity},
  Phys.\ Rept.\  {\bf 323} (2000) 183
  [arXiv:hep-th/9905111].

\bibitem{confining}
  I.~R.~Klebanov and M.~J.~Strassler,
{\it Supergravity and a confining gauge theory: Duality cascades and
  $\chi$SB-resolution of naked singularities},
  JHEP {\bf 0008}, 052 (2000)
  [arXiv:hep-th/0007191].


  J.~M.~Maldacena and C.~Nunez,
{\it Towards the large N limit of pure N = 1 super Yang Mills}, 
  Phys.\ Rev.\ Lett.\  {\bf 86}, 588 (2001)
  [arXiv:hep-th/0008001].


  
  \bibitem{KStech}
  
  K.~Tod,
  {\it All Metrics Admitting Supercovariantly Constant Spinors},
  Phys.\ Lett.\ B {\bf 121} (1983) 241.

  J.~P.~Gauntlett, J.~B.~Gutowski, C.~M.~Hull, S.~Pakis and H.~S.~Reall,
  {\em All supersymmetric solutions of minimal supergravity in five dimensions,}
  Class.\ Quant.\ Grav.\  {\bf 20} (2003) 4587
  [arXiv:hep-th/0209114].

  J.~P.~Gauntlett and J.~B.~Gutowski,
{\it All supersymmetric solutions of minimal gauged supergravity in five
  dimensions},
  Phys.\ Rev.\ D {\bf 68}, 105009 (2003)
  [Erratum-ibid.\ D {\bf 70}, 089901 (2004)]
  [arXiv:hep-th/0304064].




  D.~Martelli and J.~Sparks,
{\em G-structures, fluxes and calibrations in M-theory,}
  Phys.\ Rev.\ D {\bf 68} (2003) 085014
  [arXiv:hep-th/0306225].
  
  J.~B.~Gutowski, D.~Martelli and H.~S.~Reall,
{\it All supersymmetric solutions of minimal supergravity in six dimensions},
  Class.\ Quant.\ Grav.\  {\bf 20}, 5049 (2003)
  [arXiv:hep-th/0306235].
  
  M.~M.~Caldarelli and D.~Klemm,
  {\it All supersymmetric solutions of N = 2, D = 4 gauged supergravity},
  JHEP {\bf 0309} (2003) 019
  [arXiv:hep-th/0307022].
  
  U.~Gran, J.~Gutowski and G.~Papadopoulos,
  {\it The spinorial geometry of supersymmetric IIB backgrounds},
  Class.\ Quant.\ Grav.\  {\bf 22}, 2453 (2005)
  [arXiv:hep-th/0501177].
  
  J.~B.~Gutowski and W.~Sabra,
{\it General supersymmetric solutions of five-dimensional supergravity},
  JHEP {\bf 0510} (2005) 039
  [arXiv:hep-th/0505185].
  
  U.~Gran, J.~Gutowski, G.~Papadopoulos and D.~Roest,
  {\it Systematics Of IIB Spinorial Geometry},
  Class.\ Quant.\ Grav.\  {\bf 23}, 1617 (2006)
  [arXiv:hep-th/0507087].



 
\bibitem{Nkim05}
  N. Kim,
  \textit{$AdS_3$ solutions of IIB supergravity from D3-branes}.
  JHEP {\bf 0601} (2006) 094,
  [hep-th/0511029]


\bibitem{m2}

  J.~P.~Gauntlett, N.~Kim, S.~Pakis and D.~Waldram,
{\it Membranes wrapped on holomorphic curves},
  Phys.\ Rev.\ D {\bf 65} (2002) 026003
  [arXiv:hep-th/0105250].

  T.~Z.~Husain, {\it M2-branes wrapped on holomorphic curves},
  JHEP {\bf 0312}, 037 (2003)
  [arXiv:hep-th/0211030].






  
\bibitem{Kim:2006ag}
  N.~Kim,
   {\em Sasaki-Einstein manifolds and their spinorial geometry,}
  %
  J.\ Korean Phys.\ Soc.\  {\bf 48} (2006) 197.

\bibitem{Lin:2004nb}
  H.~Lin, O.~Lunin and J.~Maldacena,
  {\em Bubbling AdS space and 1/2 BPS geometries,}
  JHEP {\bf 0410} (2004) 025
  [arXiv:hep-th/0409174].


\bibitem{Gauntlett:2006af}
  J.~P.~Gauntlett, O.~A.~P.~Mac Conamhna, T.~Mateos and D.~Waldram,
{\it Supersymmetric AdS(3) solutions of type IIB supergravity},
  arXiv:hep-th/0606221.

\bibitem{Gauntlett:2004zh}
  J.~P.~Gauntlett, D.~Martelli, J.~Sparks and D.~Waldram,
  {\em Supersymmetric AdS(5) solutions of M-theory,}
  Class.\ Quant.\ Grav.\  {\bf 21} (2004) 4335
  [arXiv:hep-th/0402153].
  
  J.~P.~Gauntlett, D.~Martelli, J.~Sparks and D.~Waldram,
  {\em Sasaki-Einstein metrics on S(2) x S(3),}
  Adv.\ Theor.\ Math.\ Phys.\  {\bf 8} (2004) 711
  [arXiv:hep-th/0403002].

  





\bibitem{Gauntlett:2002fz}
  J.~P.~Gauntlett and S.~Pakis,
{\em The geometry of D = 11 Killing spinors},
  JHEP {\bf 0304} (2003) 039
  [arXiv:hep-th/0212008].
  
  J.~P.~Gauntlett, J.~B.~Gutowski and S.~Pakis,
{\em The geometry of D = 11 null Killing spinors},
  JHEP {\bf 0312}, 049 (2003)
  [arXiv:hep-th/0311112].


\bibitem{Donos:2006iy}
  A.~Donos,
{\it A Description Of 1/4 Bps Configurations In Minimal Type IIB SUGRA},
  arXiv:hep-th/0606199.




\end{thebibliography}
\end{document}